\documentclass[aps,prl,twocolumn,superscriptaddress]{revtex4-2}

\usepackage{xcolor,graphicx}
\usepackage{amsmath,amssymb,mathtools}
\usepackage[english]{babel}

\makeatletter
\let\ORIbbl@fixname\bbl@fixname
\def\bbl@fixname#1{%
  \@ifundefined{languagealias@\expandafter\string#1}
    {\ORIbbl@fixname#1}
    {\edef\languagename{\@nameuse{languagealias@#1}}}%
}
\newcommand{\definelanguagealias}[2]{%
  \@namedef{languagealias@#1}{#2}%
}
\makeatother

\definelanguagealias{EN}{english}
\definelanguagealias{en}{english}

\begin{document}

\title{Decomposing large unitaries into multimode devices of arbitrary size}
\author{Christian Arends}
	\affiliation{Department of Mathematics, Aarhus University, Ny Munkegade 118, 8000 Aarhus C, Denmark}
	\affiliation{PhoQS, Universit\"at Paderborn, Warburger Strasse 100, 33098 Paderborn, Germany}
\author{Lasse Wolf}
	\affiliation{Institute of Mathematics, Universit\"at Paderborn, Warburger Strasse 100, 33098 Paderborn, Germany}
\author{Jasmin Meinecke}
	\affiliation{Max-Planck-Institut f\"{u}r Quantenoptik, Garching, Germany}
\affiliation{
Department f\"{u}r Physik, Ludwig-Maximilians-Universit\"{a}t, M\"{u}nchen, Germany}
\affiliation{
Munich Center for Quantum Science and Technology (MCQST), M\"{u}nchen, Germany}
\affiliation{
Institute of Solid State Physics, Technische Universität Berlin, 10623 Berlin, Germany
}
\author{Sonja Barkhofen}
	\affiliation{PhoQS, Universit\"at Paderborn, Warburger Strasse 100, 33098 Paderborn, Germany}
	\affiliation{Department of Physics, Universit\"at Paderborn, Warburger Strasse 100, 33098 Paderborn, Germany}
\author{Tobias Weich}
	\affiliation{Institute of Mathematics, Universit\"at Paderborn, Warburger Strasse 100, 33098 Paderborn, Germany}
	\affiliation{PhoQS, Universit\"at Paderborn, Warburger Strasse 100, 33098 Paderborn, Germany}
\author{Tim J. Bartley}
		\affiliation{PhoQS, Universit\"at Paderborn, Warburger Strasse 100, 33098 Paderborn, Germany}
		\affiliation{Department of Physics, Universit\"at Paderborn, Warburger Strasse 100, 33098 Paderborn, Germany}

\date{\today}

\begin{abstract}
Decomposing complex unitary evolution into a series of constituent components is a cornerstone of practical quantum information processing. While the decompostion of an $n\times n$ unitary into a series of $2\times2$ subunitaries is well established (i.e.\@ beamsplitters and phase shifters in linear optics), we show how this decomposition can be generalised into a series of $m\times m$ multimode devices, where $m>2$. If the cost associated with building each $m\times m$ multimode device is less than constructing with $\frac{m(m-1)}{2}$ individual $2\times 2$ devices, we show that the decomposition of large unitaries into $m\times m$ submatrices is is more resource efficient and exhibits a higher tolerance to errors, than its $2\times 2$ counterpart. This allows larger-scale unitaries to be constructed with lower errors, which is necessary for various tasks, not least Boson sampling, the quantum Fourier transform and quantum simulations.
\end{abstract}

\maketitle

Unitary transformations are the basis of quantum information processing and quantum simulation. While transformations on a small number of modes are relatively straight forward, many algorithms and applications require the implementation of joint unitary transformations on a large number of modes; pertinent examples include Boson sampling~\cite{aaronson_computational_2011,hamilton_gaussian_2017}, the quantum Fourier transform~\cite{shor_algorithms_1994,coppersmith_approximate_2002},  quantum photonic simulation~\cite{aspuru-guzik_photonic_2012} and, outside of quantum photonics, neuromorphic computing~\cite{shastri_photonics_2021}. Typically these large $n\times n$ unitaries  are constructed from a decomposition into a collection of smaller $2\times 2$ unitaries. While this approach is used independent of the physical platform, (see {e.g.}~\cite{ramakrishna_explicit_2000})  it is very common in linear optics where $2\times 2$ transformations can be easily implemented by beam splitters and phase shifters,  the established building blocks of linear optics. In their seminal paper, Reck et al. demonstrated how to mathematically decompose any $n\times n$ unitary in a product of $n\left(n-1\right)/2$ $2\times 2$ subunitaries. Their decomposition results in
a triangular array of beam splitters and phase shifters, which can be programmed to implement an arbitrary linear transform of optical modes~\cite{reck1994experimental}. This scheme has been refined by Clements et al. enhancing the loss tolerance using a symmetric arrangement of $2\times 2$ splitters~\cite{clements_optimal_2016}. Furthermore, the length of the circuit could be shortened with symmetric $2\times2$ splitters~\cite{bell_further_2021}. Indeed, with the advent of integrated optics, in which many beam splitters and phase shifters can be implemented on a small footprint, large and complex unitary operations have been demonstrated using this approach, not least linear optics quantum computing~\cite{carolan_universal_2015,
bogaerts_programmable_2020,harris_large-scale_2016}, Boson sampling~\cite{spring_boson_2013,
crespi_integrated_2013,
madsen_quantum_2022}, quantum simulation~\cite{sparrow_simulating_2018, van_der_meer_experimental_2023,somhorst_quantum_2023}, and neuromophic computing~\cite{shainline_superconducting_2017,feldmann_all-optical_2019}. Nevertheless, while sources and detectors scale linearly with the network dimension $n$, the required number of beam splitters and phase shifters scales with $\mathcal{O}(n^2)$, in order to implement an arbitrary $n\times n$ unitary. It is therefore pertinent to investigate building large unitaries starting from larger building blocks.

Beyond decompositions of large unitaries into arrays of beam splitters, other approaches such as multiport integrated devices and $3\times 3$ fiber tritters have also been investigated~\cite{soldano_optical_1995} in the context of quantum interference~\cite{weihs_two-photon_1996,peruzzo_multimode_2011,spagnolo_three-photon_2013,menssen_distinguishability_2017}. Higher-order modal manipulation of quantum light has also been investigated beyond the spatial degree of freedom, which is highly promising for experimentally implementing unitaries of larger size. For example, methods for manipulating orbital angular momentum modes have shown up to seven modes~\cite{mirhosseini_high-dimensional_2015}, however generalised manipulation remains challenging~\cite{babazadeh_high-dimensional_2017}. In the frequency degree of freedom, operations on 10 modes have been shown~\cite{kues_-chip_2017}, while operations on hybrid time-frequency modes have also been demonstrated up to 64 modes~\cite{brecht_photon_2015,ansari_tailoring_2018, de2022measurement, serino2023realization}. This begs the question: how can larger $n\times n$ unitaries be constructed from $m\times m$ constituent unitaries, where $m>2$?

The answer to this question becomes practically relevant only when the $m\times m$ constituent device outperforms (by some reasonable metric) its own decomposition into $2\times 2$ components. In other words, if the performance cost ({e.g.} loss, fidelity, production cost, etc) associated with producing a $m\times m$ unitary is greater than the $\frac{m(m-1)}{2}$ different $2\times2$-unitaries, the $2\times 2$ decomposition is more efficient.
Nevertheless, once it becomes cheaper
 to directly fabricate a device realizing an $m\times m$ unitary compared to building it out of $2\times2$ phase shifter/beam splitter cascades, the question how one efficiently can build an $n\times n$ unitary from its  $m\times m$ subunitaries is of the utmost relevance. Indeed, in all the aforementioned physical implementations, the number of possible modes is physically restricted far below the size the desired matrix size for quantum computing applications.

In this paper, we generalise unitary decomposition of an $n\times n$ unitary into $m\times m$ submatrices, where $n>m\geq2$. This provides a significant scaling advantage whenever it is cheaper to produce a $m\times m$ unitary directly, compared to building it out of $2\times 2$ unitaries. We provide an algorithmic approach to find this decomposition, which uses the best-known minimum number of submatrices. We also show that quality thresholds exist when comparing larger devices to the established beam splitter decomposition. This provides a route to implementing large scale devices, which, by reducing the total number of components, are more tolerant to the errors caused by the components individually.

The algorithm to decompose a unitary $n\times n-$matrix $U$
into a product of smaller unitary matrices of dimension $m\times m$ runs as follows:  The key task is to find  $m\times
m-$matrices $\tilde Q_1,\ldots,\tilde Q_N$ (properly embedded as $n\times n-$matrices)
such that $U\tilde Q_1\ldots \tilde Q_N$ is an upper triangular matrix. Note that  any unitary upper triangular matrix is automatically  a diagonal unitary matrice $D$. Such diagonal matrices are experimentally easy to realize because it consists only of a phase shift in each individual mode. Summarizing, our algorithm will allow to write the large unitary $U$ as $U= D \tilde Q_N^{-1}\dots \tilde Q_1^{-1}$ thus we have factorized $U$ into $m\times m$-unitaries and  final phase shifts.

In order to achieve upper triangular matrices we
use the RQ-decomposition which, for any $m\times m-$matrix $A$,
ensures the existence of a unitary matrix $Q$ and an upper triangular matrix
$R$ (i.e.\@ $R_{ij}=0$ for $i>j$) such that $A=RQ^{-1}$ (see e.g.\@ \cite[Section~5.2]{golub1996a}). In particular, $AQ$ is
upper triangular so that we can transform any matrix into an upper triangular
one by right multiplication with a unitary.

\begin{figure*}
        \centering
        \includegraphics[height=4.3cm]{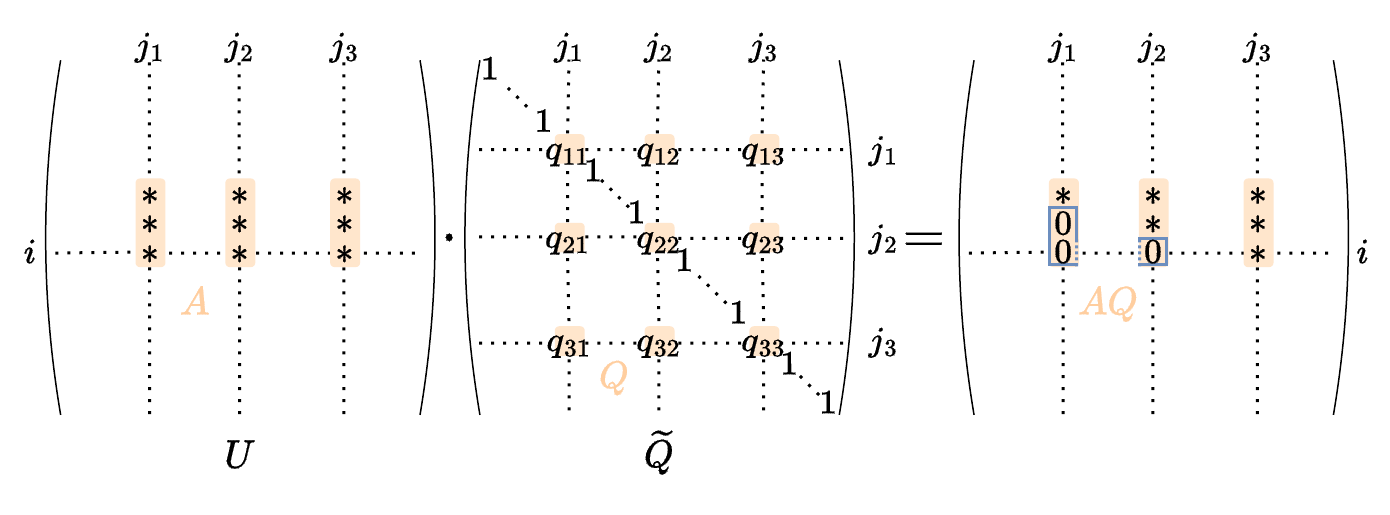}\hspace{0.1cm}
	\includegraphics[height=4.3cm]{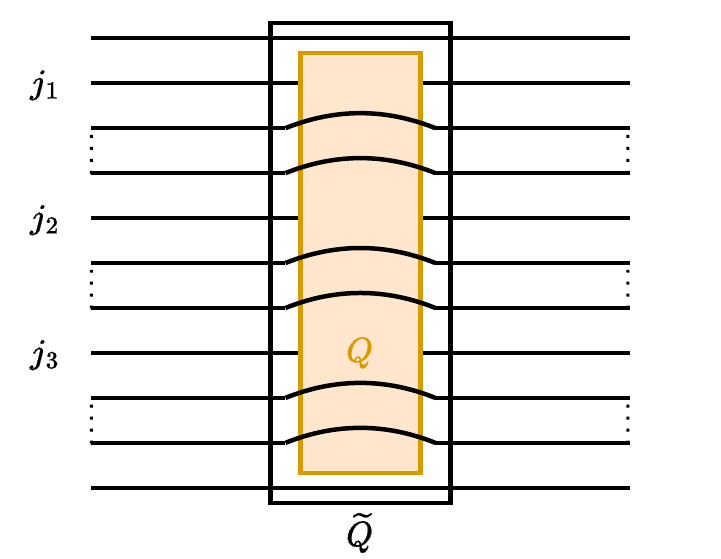}
        \caption{Example for $m=3$. The matrix $\tilde Q$ only affects the input modes $j_1,\ldots, j_m$.}
        \label{fig:matrix_embedding}
\end{figure*}

We now describe how to use the
RQ-decomposition to create zeros at predefined positions in a large unitary
matrix $U$. We will use this multiple times to create zeros at all places below the diagonal.
Let us fix $m$ columns $1\leq j_1<\ldots<j_m\leq n$ and a base row
$i\in\{m,\ldots,n\}$. This choice gives rise to a $m\times m$-matrix $A$
consisting of the entries of $U$ which are contained in the columns
$j_1,\ldots,j_m$ and in the rows $i-m+1,\ldots,i$ (see Fig.~\ref{fig:matrix_embedding} for an illustration of this embedding for $m=3$).
Our goal is to transform this matrix into an upper triangular form. First, by
the RQ-decomposition, there is a unitary matrix $Q$ such that $AQ=R$ is upper
triangular. We now show how to properly embed the matrix $Q$ into an $(n\times
n)-$matrix such that we can create zeros in our original matrix $U$. For this
let $q_{k\ell}$ denote the entries ofthe $m\times m$ matrix $Q$ and build an $n\times n$-matrix $\tilde{Q}$ with entries $\tilde{q}_{k\ell}$ as follows: Start with the  identity matrix and set $\tilde{q}_{j_k, j_{\ell}} \coloneqq q_{k \ell}$, i.e.\@ we embed $Q$ into the identity matrix
at the $m\times m$-submatrix given by the entries having their row and column
coordinates both in $\{j_1 , \ldots , j_m\}$.

Thus, $U\tilde Q$ has an upper triangular ${m\times m}$-sub\-ma\-trix.
More precisely, the entries in row $l$ and column $j_k$ are zero for $l=i-m+1+k,\ldots , i$ and $k=1,\ldots, m$.
Note
that the multiplication with $\tilde Q$ from the right only affects the columns
$j_1,\ldots,j_m$ of $U$. Moreover, if there are only zeros in these columns
below the $i$th row, all these zeros are maintained by this multiplication.
To state this differently, if $U$ has zeros at $(l,j_k)$, $k=1,\ldots,m$, this is equivalent to saying that an input state that occupies only the modes $j_1,\ldots, j_m$ is transformed to an output state where the $l$th mode is not occupied.
Since $\tilde Q$ only affects the input modes $j_1,\ldots,j_m$ this property is preserved if we apply the matrix $\tilde Q$ before applying $U$ which means that we consider $U\tilde Q$.

\begin{figure}
   \includegraphics[width=0.45\textwidth]{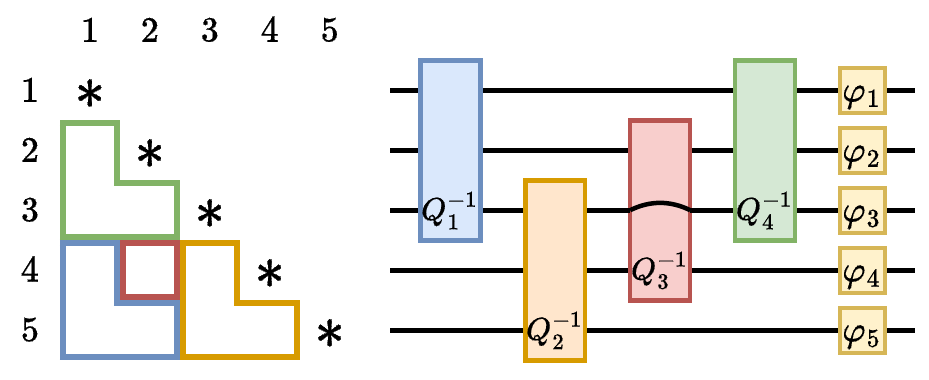}
   \caption{Decomposition for $m=3$ and $n=5$ with experimental setup. Here, the $\varphi_i$ stand for the phase shifts which together form the diagonal matrix $D$.}
    \label{fig:experimental_setup}
\end{figure}

The algorithm to find the matrices $Q_i$ we are presenting works
similarly to a tetris game: In each step we create zeros in a
specific row by inserting blocks of zeros using the technique previously
described.
If $m=2$ then this algorithm is exactly the algorithm of \cite{reck1994experimental}.
We start by creating zeros in the bottom row. By selecting the
first $m$ columns and using the RQ-decomposition as above, we can build an
upper triangular $m\times m$-block in the lower left corner of $U$ (see Fig.\@ \ref{fig:experimental_setup}).
This gives us the matrix $\tilde Q_1$. We now proceed in the same way with the next $m$ columns without zero entries to create
more and more triangle-shaped blocks of zeros in that row until the number $m'$ of remaining
non-zero entries is less that $m$ and use (potentially) one more matrix of size $m'$ to
fill in the remaining zeros -- a triangle-shaped block of size $m'$. As we already
created some zeros in the penultimate row, we only have to create new zeros
at the places which are not already covered.
Here we have to split up the triangle-shaped block and use selected columns $j_1,\ldots, j_m$ as described in the previous paragraph.
In general, the algorithm is
structured as follows: \begin{enumerate}
	\item Let $i\in \{2,\ldots,n\}$ be the smallest value such that in the rows $i+1,\ldots,n$ all entries below the diagonal are zero.
		In the first step described above we generically have no zeros in the last row, i.e.\@ $i=n$.
        \item Consider the $i$th row and pick the first $m$ non-zero entries in that row which are on the left hand side of the diagonal or on the diagonal. Denote the corresponding columns by $j_1,\ldots,j_m$. If there are just $2\leq m'<m$ non-zero entries left in that row, proceed with $m'$ instead of $m$.
        \item Choose a unitary matrix corresponding to the row $i$ and the columns $j_1,\ldots,j_m$ to create an upper triangular block of size $m$, as described above.
        \item Repeat until $U$ is transformed into an upper triangular matrix.
\end{enumerate}

\begin{figure}
        \centering
        \includegraphics[width=0.7\linewidth]{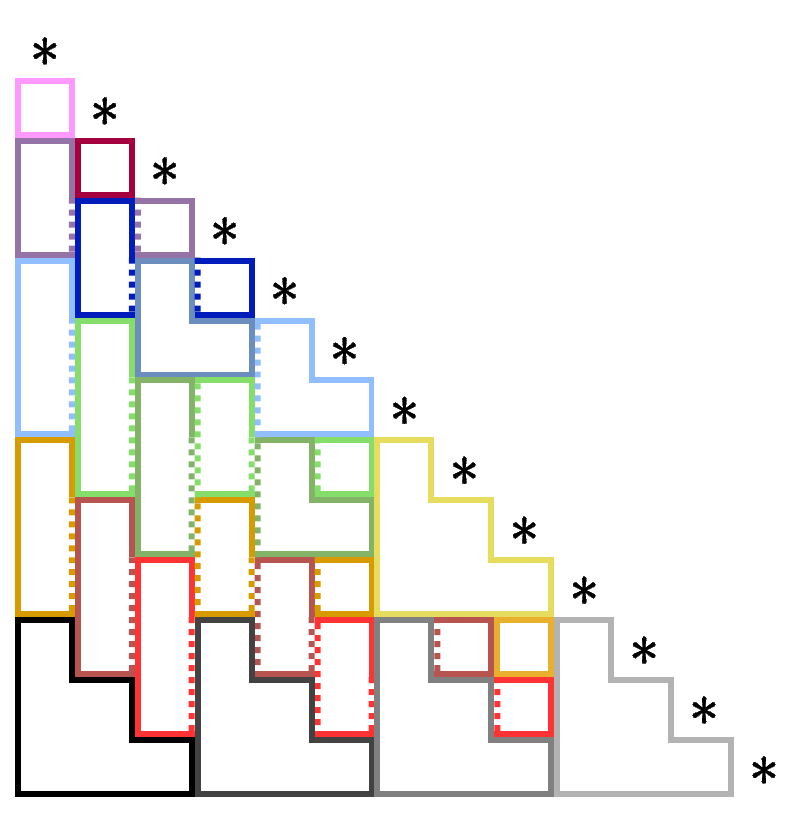}
        \caption{Illustration of the algorithm using triangle blocks for $m=4$ and $n=13$.}
	\label{fig:tetris4}
\end{figure}
\begin{figure}
	\centering
	\includegraphics[width=0.95\linewidth]{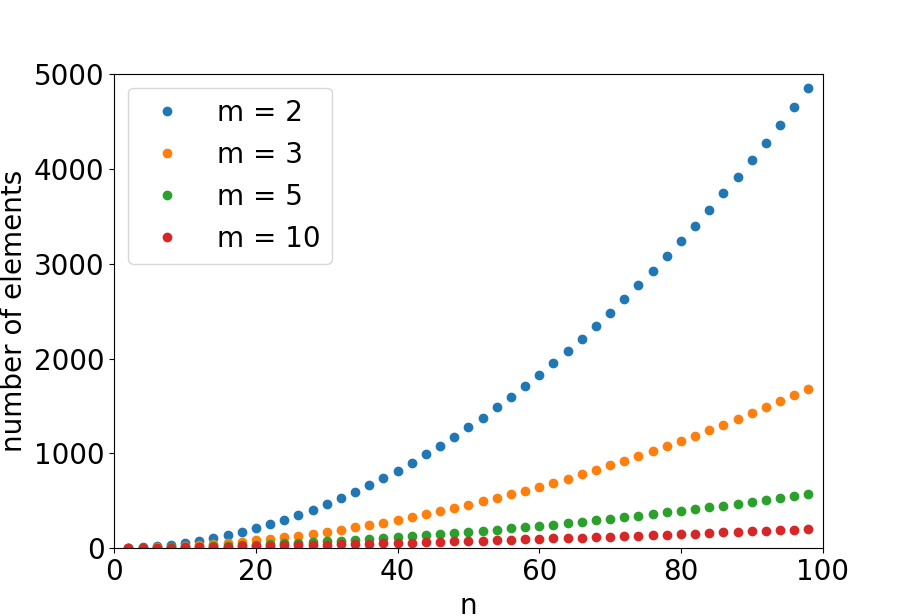}
	\caption{Scaling behaviour of our algorithm for the number of elements to construct a unitary of size $n$ according to $\frac{n(n-1)}{m(m-1)}+n-1$}
	\label{fig:elements_scaling}
\end{figure}

Each $m\times m$-submatrix $\tilde Q_i$ creates $\frac 12 (m-1)m$ zeros in the matrix $U$.
In total we have to create $\frac 12 (n-1)n$ zeros to end up with a diagonal matrix.
Hence, we need at most $\frac{n(n-1)}{m(m-1)}$ matrices $\tilde Q_i$ which come from a $m\times m$-matrix.
In addition, our algorithm requires at most one $m'\times m'$-submatrix with $m'<m$ for each row.
Hence, we end up with at most $\frac{n(n-1)}{m(m-1)}+n-1$ matrices $\tilde Q_i$, see Fig.~\ref{fig:elements_scaling}.

If $c_m$ is the relevant cost of a $m\times m$ unitary then the total cost for constructing the $n\times n$ matrix out of $m\times m$ matrices is $C_{n,m} \leq c_m(\frac{n(n-1)}{m(m-1)}+n-1)$  (under the assumption that the total cost is linear in the number of utilized devices). Recall that the Reck or Clements scheme requires $\frac{1}{2}n(n-1)$ different $2\times 2$-unitaries which leads to a relevant cost of $C_{n,2}=\frac{c_2}{2}n(n-1)$ for the realization of a $2\times 2$ unitary. Comparing these two cost function one sees directly that whenever $c_m < \frac{1}{2}m(m-1)$ (i.e. the $m\times m$ device is cheaper then building the matrix out of $2\times 2$ matrices) one sees that for large $n$, $C_{n,m}<C_{n,2}$, i.e. it is advantageous to build the $n\times n$ matrix out of $m\times m$ instead of the traditional $2\times 2$ beam splitter phase shifter cascades.

\begin{figure}
	\centering
	\includegraphics[width=\linewidth]{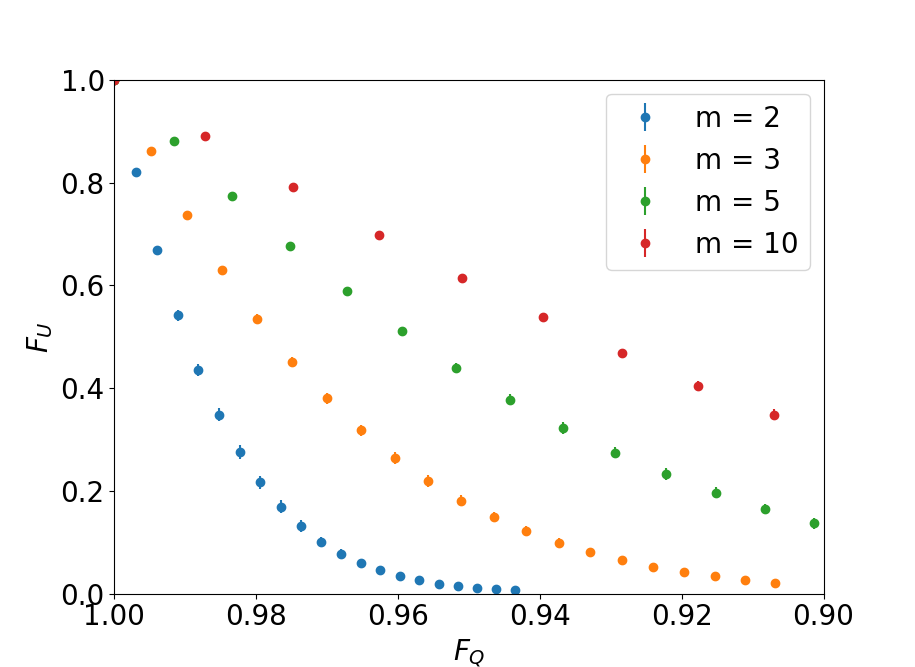}
	\caption{Numerical simulations of the fidelity for $n = 50$. For each data point we reconstructed 100 random unitary matrices $U$, each with 20 different perturbations and averaged over all 2000 reconstructions. The size of the calculated submatrices is indicated in legend. The error bars of the statistical fluctuations are smaller than the symbol size}
	\label{fig:numerics}
\end{figure}

\begin{figure}
	\centering
	\includegraphics[width=\linewidth]{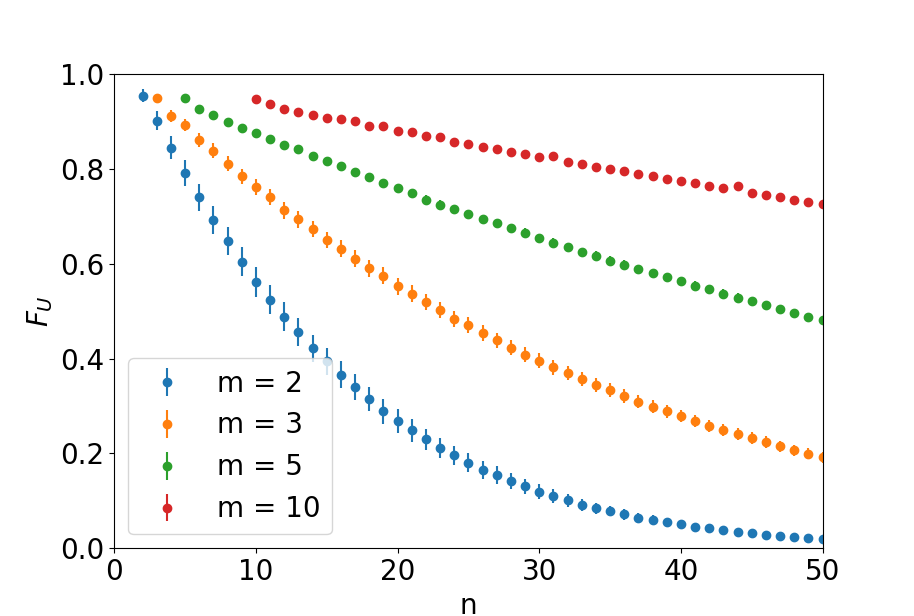}
	\caption{Numerical simulations of the fidelity for increasing $n \ge m$ at a fixed component quality $F_Q = 0.95\pm 0.0005$. For each data point we reconstructed 100 random unitary matrices $U$, each with 20 different perturbations. The size of the submatrices is indicated in legend. The error bars of the statistical fluctuations are smaller than the symbol size.
	}
	\label{fig:numericsNs}
\end{figure}

We numerically implemented the above described algorithm in order to test its performance as well as its robustness against perturbations:
For this, we randomly chose a $n\times n$-unitary matrix $U$ and decomposed it into $m\times m$-unitaries $Q_i$ as described above. We then perturbed the calculated $m\times m$-submatrices $Q_i$ by adding random numbers from a normal distribution with specified width (``noise strength'') to real and imaginary part of the entries, respectively.
As a figure of merit for the strength of the perturbation we consider the fidelity of the perturbed submatrices $Q_\mathrm{pert}$ and the original submatrices $Q$ given by
\begin{eqnarray}
	 F (Q, Q_\mathrm{pert}) := \left|\frac{\frac{1}{m}\mathrm{Tr}(Q^\dag \cdot Q_\mathrm{pert})}{\sqrt{\frac{1}{m}\mathrm{Tr}( Q_\mathrm{pert}^\dag \cdot  Q_\mathrm{pert})}}\right|^2 ,
\end{eqnarray}
This will of course vary for any realized perturbation so we take the expected reconstruction fidelity $F_Q := \mathbb E(F (Q, Q_\mathrm{pert}))$ as a measure for the precision of our individual components $Q$.
Next, we reconstruct a $(n\times n)$ matrix $U_\mathrm{pert}$ from the $\tilde{Q}_\mathrm{pert}$.
The final metric to analyse the robustness of $U$ is defined by $F (U, U_\mathrm{pert})$ respectively by the expected fidelity $F_U:=\mathbb E(F(U,U_\mathrm{pert}))$ and plotted in Figs. \ref{fig:numerics} and \ref{fig:numericsNs}.
In the first figure, we fixed the matrix size at $n = 50$ and plot the dependence of the reconstruction fidelity $F_U$ as a function of the component quality $F_Q$ for different submatrix size $m = 2,3,5,10$.
It can be clearly seen from the figure that the reconstruction fidelity of $F_U$ drops quickly with component quality $F_Q$.
The smaller the submatrices are (i.e. the smaller $m$) the steeper is the fidelity drop.
Thus, already the increase in component size from $m = 2$ to $m= 3$ improves the fidelity of the final unitary significantly and proves a much higher robustness of the reconstructed matrix $U_\mathrm{pert}$.
Second, we analysed the question, how big can the final system size (i.e. unitary size $n$) become, given components of size $m$ achieving a specified quality $F_Q$.
The results for a fixed value $F_Q = 0.95\pm 0.0005$ and $m = 2,3,5,10$ are presented in Fig. \ref{fig:numericsNs}.
Again, we observe already for $m = 3$ a significant advancement in fidelity which enables the realisation of much bigger matrices, i.e. quantum networks, at reasonable fidelities.
The advantage increases with the size of the submatrices, as expected.

In conclusion, we have presented an algorithm to decompose large $n\times n$-unitaries into smaller constituent $m\times m$-subunitaries. We have shown that these become  more tolerant to loss and errors as $m$ increases, yielding the intuition that larger unitaries are more effectively building from larger building blocks. This has implications for building large-scale unitary dynamics on quantum systems, in particular linear optics, where the decomposition in terms of $2\times 2$ beam splitters and phase shifters has become ubiquitous. Exploring other devices which intrinsically operate on a larger set of modes simultaneously is thus highly advantageous, and may simplify the path towards practical large scale devices.

\subsection*{Acknowledgements}
	This work has received funding from the European Union's Horizon 2020 research and innovation program under grant agreement No 665148 as well as from Deutsche Forschungsgemeinschaft (DFG) Grant No. WE 6173/1-1 (Emmy Noether group “Microlocal Methods for Hyperbolic Dynamics”) and SFB-TRR 358/1 2023 — 491392403 (CRC ``Integral Structures in Geometry and Representation Theory'')
	as well as funding by the Ministerium für Kultur und Wissenschaft des Landes Nordrhein-Westfalen via the project PhoQC. JDAM acknowledges support by the DFG under Germany’s Excellence Strategy EXC-2111 390814868.

%

\end{document}